\documentclass[seceq,supplement]{ptptex}

\usepackage{graphicx}
\usepackage{wrapft}

\newcommand{\psibar}{\overline{\psi}}
\newcommand{\chibar}{\overline{\chi}}
\newcommand{\phibar}{\overline{\phi}}
\newcommand{\Xbar}{\overline{X}}
\newcommand{\jbar}{\overline{\jmath}}
\newcommand{\gf}{\gamma_5}
\newcommand{\cgf}{C\gamma_5}
\newcommand{\Mt}{\widetilde{M}}

\preprintnumber[5.2cm]{TRINLAT-03/09,SWAT/03/089,BI-TP 2003/34}

\title{%
Lattice simulations of 2-colour QCD with Wilson fermions
}

\author{
Jon-Ivar \textsc{Skullerud}$^{1,2,}$\footnote{Talk by Jon-Ivar Skullerud},
Shinji \textsc{Ejiri}$^3$,
Simon \textsc{Hands}$^4$ and
Luigi \textsc{Scorzato}$^{5,6}$%
}

\inst{%
$^1$ITF, Universiteit van Amsterdam, Valckenierstraat 65, 1018 XE
  Amsterdam,\\ The Netherlands\\
$^2$School of Mathematics, Trinity College, Dublin 2, Ireland\\
$^3$Fakult\"at f\"ur Physik, Universit\"at Bielefeld,
  Universit\"atsstra{\ss}e 25, 33615 Bielefeld, Germany\\ 
$^4$Department of Physics, University of Wales Swansea, Singleton
  Park,\\ Swansea SA2 8PP, Wales\\
$^5$Theory Group, DESY, Notkestra{\ss}e 85, 22603 Hamburg, Germany\\
$^6$Institut f\"ur Physik, Humboldt Universit\"at zu Berlin, 12489
  Berlin, Germany%
}

\abst{%
We report on the status of our simulations of two-colour QCD with two
flavours of Wilson quarks, at nonzero chemical potential and diquark
source.  Preliminary results are presented for the static
quark potential, gluon propagator, and meson and diquark spectrum.
}

\begin{document}

\maketitle

\section{Motivation}

QCD at non-zero density is currently an area of great activity, as the
wide range of presentations at this workshop bears witness.  What
is missing at present, however, is a first-principles, nonperturbative
approach to QCD at high density and low temperature, without which we
have little hope of obtaining quantitative predictions regarding the
expected complex phase structure in this regime.  Lattice QCD would be
the natural candidate for such an approach, but it is severely
hampered by the fact that the euclidean action becomes complex when a
non-zero chemical potential is introduced.  Ref.~\citen{Muroya:2003qs}
gives an overview of lattice QCD at non-zero density and the problems
encountered.

We may avoid these problems by studying model theories which share
some of the salient features of QCD, while having a positive definite
fermion determinant also for non-zero chemical potential.  Examples of
such theories are 2-colour QCD\cite{Hands:1999md}, QCD with adjoint
fermions\cite{Hands:2000ei}, NJL
models\cite{Hands:2002mr,Walters:2003nn}, and QCD with isospin
chemical potential\cite{Kogut:2002zg}.  Here we will be concentrating
on 2-colour QCD, which in the gauge sector is expected to be very
similar to real QCD, although the baryon sector and chiral properties
of the two theories are radically different.

It is of fundamental importance that a theory to be studied using
lattice regularisation have a well-defined continuum limit.  In the
context of non-abelian gauge theories, this implies that the theory
should exhibit asymptotic freedom.  For SU(2) the second term in the
perturbative $\beta$-function changes sign for $N_f\geq5.6$, with
radical consequences for the infrared behaviour of the theory;
nonperturbative effects reducing this critical value of $N_f$ still
further cannot be ruled out.  Consequently we prefer to work with
$N_f=2$ so that both asymptotic freedom (and hence a continuum limit)
and confining behavour in the infrared are guaranteed.  Most
simulations up to now have been performed using staggered fermions,
where either $N_f$ is a multiple of 4, or a square-root of the fermion
determinant is taken, making the action non-local.  To avoid the
potential pitfalls this creates we are here using the Wilson action
with $N_f=2$ quark flavours.  This action has also been employed by
the Hiroshima group \cite{Muroya:2002ry,Muroya:2003zz}, albeit with a
different gauge action, so our results may be directly compared to
theirs.

It is also worth pointing out that for $\kappa=\kappa_c$ Wilson
fermions, unlike staggered fermions, have the same global symmetry
properties as continuum 2-color QCD, as encoded by the Dyson index
$\beta=1$ \cite{Hands:2000ei}, which has consequences for the physical
spectrum of the model as analysed in Ref.~\citen{Kogut:2000ek}.

We now list some physical motivations for studying this model.

\subsection{Phase diagram}

There are indications, from studies at nonzero isospin-chemical
potential, that the curvature of the phase transition line in the
$(T,\mu)$-plane for low $\mu$ is the same as in real QCD.  If this can
be confirmed, we may gain insight into the phase diagram of real QCD
by studying 2-colour QCD, where numerical studies are not restricted
to small $\mu/T$.

At the other end of the phase diagram, at low $T$ and high $\mu$, a
possible tricritical point has been identified \cite{Kogut:2002cm}.
A confirmation of this would be welcome.  Another interesting issue at
low $T$ and intermediate $\mu$ is whether 2-colour QCD has a normal
matter phase, i.e.\ a phase with nonzero baryon number density and zero
diquark condensate --- or whether the finite-$\mu$ transition is
directly from the vacuum to a superfluid phase.  The available
evidence, as well as chiral perturbation theory studies, indicate the
latter, but the matter has not yet been finally settled.

\subsection{Spectrum}

Previous studies of 2-colour QCD with Wilson
fermions\cite{Muroya:2002ry,Muroya:2003zz} have indicated that the
vector meson becomes lighter with increasing chemical potential, in
line with the predictions of Brown--Rho scaling\cite{Brown:1991kk}
which have been put forward as a possible explanation for the CERES
results\cite{Agakishiev:1998au}.  Such an in-medium effect, if
confirmed, would provide a new testing ground in heavy-ion experiments
such as the one planned at GSI.  Closely related to in-medium
modification of vector meson properties is the possibility of vector
condensation, which may occur instead of or in addition to scalar
diquark condensation at high
density\cite{Langfeld:1997fi,Sannino:2001fd}.

The behaviour of the pseudoscalar isosinglet diquark at high density
is also of interest, since it shares the quantum numbers of the
$\eta^\prime$ meson but is accessible via standard lattice
spectroscopic techniques. Its becoming light as $\mu$ increases may
signal the restoration of U(1)$_A$ symmetry in dense
matter\cite{Schafer:2002yy}.

Finally, we hope to compare the actual behaviour of the pion and the
C-even and -odd scalar diquark masses, as well as the baryon density
and chiral and diquark condensates, with the predictions from chiral
perturbation theory\cite{Kogut:2000ek}.  The agreement (for the chiral
condensate in particular) is striking in the staggered adjoint
model\cite{Hands:2001ee} which is in the same global symmetry class
($\beta=1$) as the model we are studying here. At some point, of
course, chiral perturbation should cease to be accurate once $\mu$
becomes comparable with the mass of the vector diquark, and it is
important to know the range of parameters where this breakdown can be
anticipated.

\subsection{Gluodynamics}

Gluodynamics is arguably the sector where results from two-colour QCD
have most direct relevance to real QCD.  At zero chemical potential,
the gluodynamics of SU(2) and SU(3) are so similar that SU(2)
simulations have often been used as a `cheaper' way of obtaining
results relevant to QCD.  This is also where we most directly can
study effects of fermion loops, including Pauli blocking at high
density.  In contrast, all numerical results for spectrum and
condensates to date might conceivably have been obtained in the
quenched approximation.

Among the quantities of particular interest are:
\begin{itemize}
\item Polyakov line screening and static quark potential.  Previous
  simulations of SU(2) with Wilson fermions have reported signs of a
  deconfinement transition at large $\mu$\cite{Muroya:2003zz}, while
  no such transition has been observed with staggered fermions.
\item Gluon propagator.  This quantity is an essential ingredient also
  in the Dyson--Schwinger equation approach to gauge
  theories\cite{Roberts:2000aa,Alkofer:2000wg}.  These approaches can
  be applied equally to SU(2) and SU(3) gauge theories, so input from
  SU(2) lattice simulations may provide a valuable check on the
  methods and assumptions used.
\item Other quantities that may provide further insight into the phase
  transition include glueball masses\cite{Lombardo:2003uu} and the
  gluon condensate\cite{Baldo:2003id}.
\end{itemize}

\section{Algorithm}

Above the critical chemical potential $\mu_c$ for diquark
condensation, it is essential to have a diquark source.  This serves
two purposes:
\begin{enumerate}
  \item It lifts (near-)zero eigenvalues from the Goldstone modes,
  improving the condition number and avoiding critical slowing down.
\item It also enables the determination of diquark condensates and
  other anomalous propagation.
\end{enumerate}

The action, including diquark source terms, can be written
\begin{equation}
\begin{split}
S =& \psibar_1 M(\mu)\psi_1 + \psibar_2 M(\mu)\psi_2 \\
 & + j\Bigl[\psi_2^{tr}\tau_2\cgf\psi_1 -
 \psi_1^{tr}\tau_2\cgf\psi_2\Bigr]
 + \jbar\Bigl[\psibar_2\tau_2\cgf\psibar_1^{tr} -
 \psibar_1\tau_2\cgf\psibar_2^{tr}\Bigr] \, ,
\end{split}
\end{equation}
where $j,\jbar$ are sources for the diquark and antidiquark
respectively, and $M(\mu)$ is the usual Wilson-fermion matrix.  If we
introduce $\Mt(\mu)\equiv\gf M(\mu)$ and the new field variables
\begin{equation}
 \chi_1 = \psi_1\,, \qquad \chibar_1 = \psibar\gf\,, \qquad 
 \phi_2^{tr} = \psi_2^{tr}\tau_2\cgf\,, \qquad
 \phibar_2^{tr} = \tau_2C\psibar_2^{tr}\,,
\end{equation}
the action can be rewritten as
\begin{equation}
S = \begin{pmatrix}\phi_2^{tr} & \chibar_1\end{pmatrix}
 \begin{pmatrix} j & \Mt(\mu) \\ \Mt(-\mu) & -\jbar\end{pmatrix}
 \begin{pmatrix}\chi_1\\\phibar_2^{tr}\end{pmatrix}
\equiv \Xbar Q[\mu,j,\jbar]X \,.
\end{equation}
The matrix $Q$ is hermitean, and the determinant is negative definite if
the two sources are equal: $j=\jbar$, so this representation lends
itself to simulation using standard algorithms.

We will use the two-step multibosonic algorithm\cite{Montvay:1996ea},
which has been shown to be more efficient than Hybrid Monte Carlo
(HMC) in the dense phase\cite{Hands:2000ei}, and which is ergodic even
in the presence of singularities.  However, in the initial phase, we
have also performed simulations using HMC without a diquark source,
and these are the results we will quote in the following.

\section{Results}

In order to establish a baseline, we have first performed simulations
on $8^3\times16$ lattices at $\mu=0$, for several values of $\beta$
and $\kappa$, using the standard Wilson action both for gauge fields
and for fermions.  The parameters are given in table~\ref{tab:params},
along with the associated average spatial and timelike plaquette
values.  We have sampled configurations every 4 trajectories.
\begin{table}
\begin{center}
\begin{tabular}{rrlrrlll}
$\beta$ & $\kappa$ & $dt$ & $N_{\text{traj}}$ & acc
 & $\Box_s$ & $\Box_t$ & Name \\ \hline
1.7 & 0.1780 & 0.0125 & 7152 & 84\% & 0.47386(6) & 0.47367(6) & coarse (c)\\
1.8 & 0.1725 & 0.02   & 3856 & 70\% & 0.50007(8) & 0.49983(8) \\
    & 0.1740 & 0.01   & 216  & 89\% \\
    &        & 0.0125 & 272  & 80\% & 0.5056(2)  & 0.5052(2)  \\
    & 0.1750 & 0.01   & 600  & 80\% & 0.5140(3) & 0.5139(3)
  & light (l) \\
1.9 & 0.1650 & 0.02   & 2316 & 83\% & 0.52124(10) & 0.52098(10)
  & fine (f) \\ \hline
\end{tabular}
\end{center}
\caption{Hybrid Monte Carlo simulation parameters, $\mu=0$.  $dt$ is
  the HMC timestep, while `acc' denotes the acceptance rate.
  $\Box_s$ and $\Box_t$ are the average spatial and timelike plaquette
  values respectively.}
\label{tab:params}
\end{table}

\subsection{Setting the scale}

\begin{figure}
\centerline{\includegraphics[width=10cm,clip]{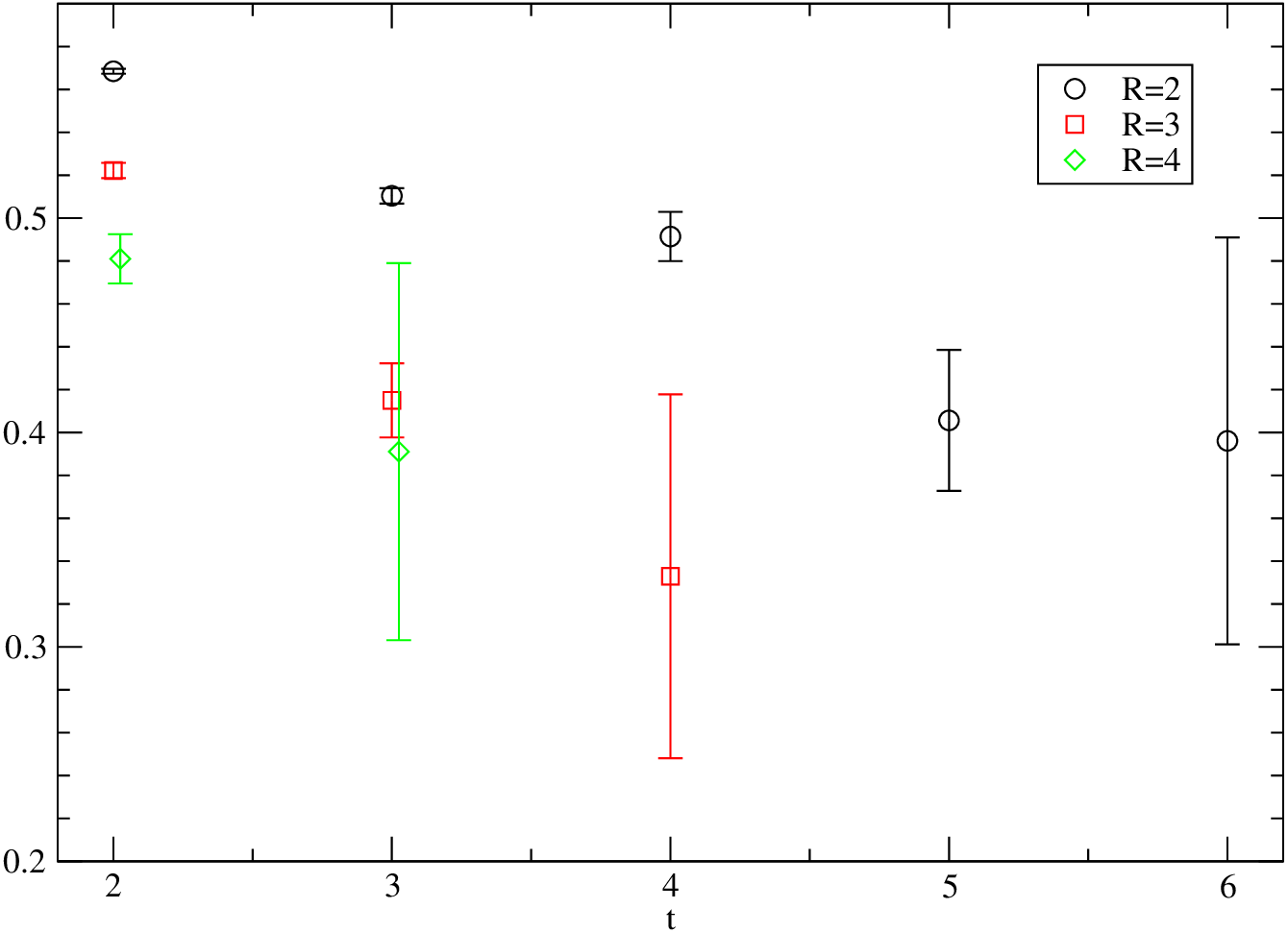}}
\caption{Creutz ratios for the coarse lattice $(\beta=1.7,
\kappa=0.178)$.}
\label{fig:creutz-b17}
\end{figure}

We have attempted to set the scale from the string tension by
computing Creutz ratios.  Figure~\ref{fig:creutz-b17} shows the
results for the coarse lattice $(\beta=1.8,\kappa=0.178)$ which is
where we have the best statistics.
It is evident from the figure that our statistics are still not
sufficient to determine the string tension from Creutz ratios with any
degree of confidence, and from the other parameter values the
situation is even worse.  Based on the trend displayed by the data, we
tentatively estimate $a^2\sigma\approx0.37(5)$.  Taking the string
tension to be $\sqrt{\sigma}\approx 420$ MeV, this gives a lattice
spacing $a\approx0.29$ fm or $a^{-1}\approx690$ MeV.

\begin{figure}
\centerline{\includegraphics[width=10cm,clip]{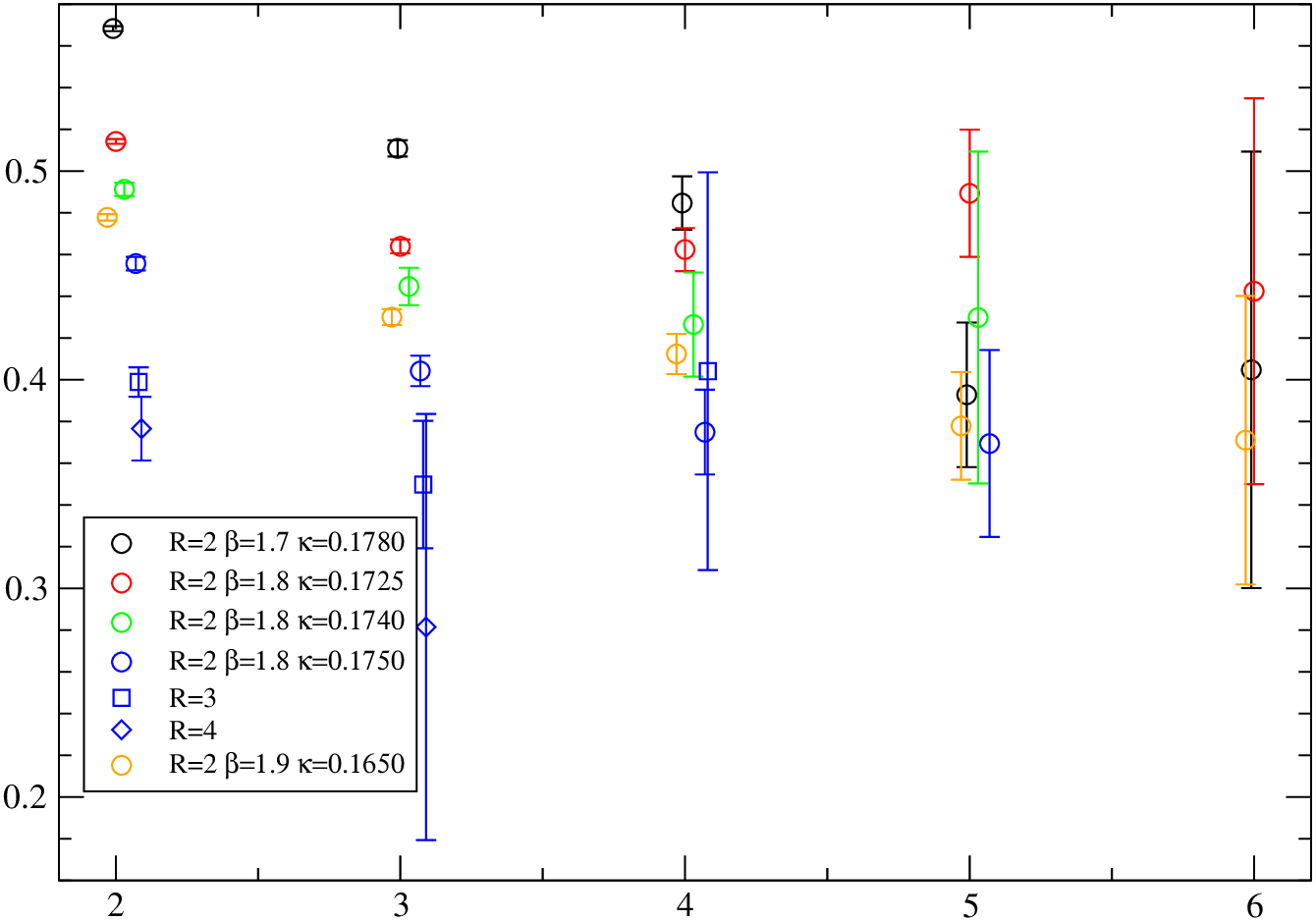}}
\caption{Creutz ratios for $R=2$, all $(\beta,\kappa)$ values.}
\label{fig:creutz-all}
\end{figure}

In Fig.~\ref{fig:creutz-all} we show Creutz ratios for $R=2$ only for
all our parameter values.  We also include the $R=3,4$ data for the
light lattice $(\beta=1.8,\kappa=0.175)$, to give an indication of the
trend.
We may attempt to determine ratios of lattice spacing by simply taking
the ratios of Creutz ratios at $(R=2,T=3)$ where the statistics permit
this.  The results are shown in table~\ref{tab:Ra}.  These
numbers are likely to be overestimates, since ratios for $R=2$
appear to increase rather than decrease with $T$.
\begin{table}
\begin{center}
\begin{tabular}{rrlrrr}
 &  & Creutz & \multicolumn{3}{c}{Gluon}\\ \hline
$\beta$ & $\kappa$ & $R_a$ & $R_a$ & $R_Z$ & $\chi^2/N_{df}$ \\ \hline
1.7 & 0.1780 & 1 & 1 & 1 & --- \\
1.8 & 0.1725 & 0.95  & 0.84(1) & 1.06 & 9.7 \\
    & 0.1740 & 0.93  & 0.76(2) & 1.06 & 2.0 \\
    & 0.1750 & 0.89  & 0.69(3) & 1.06 & 2.7 \\
1.9 & 0.1650 & 0.92  & 0.74(1) & 1.14 & 20 \\ \hline
\end{tabular}
\end{center}
\caption{Ratios of lattice spacings $R_a=a/a_c$, where $a_c$ is the
  lattice spacing on the coarse lattice.  `Creutz' denotes the
  ratios of Creutz ratios for $R=2,T=3$, while `Gluon' denotes the
  outcome of matching the gluon propagator as described in
  Ref.~\citen{Leinweber:1998uu}, with $R_Z$ being the ratio of
  renormalisation constants $Z_3/Z_3^c$.}
\label{tab:Ra}
\end{table}

\begin{figure}
\centerline{\includegraphics[width=\halftext,clip]{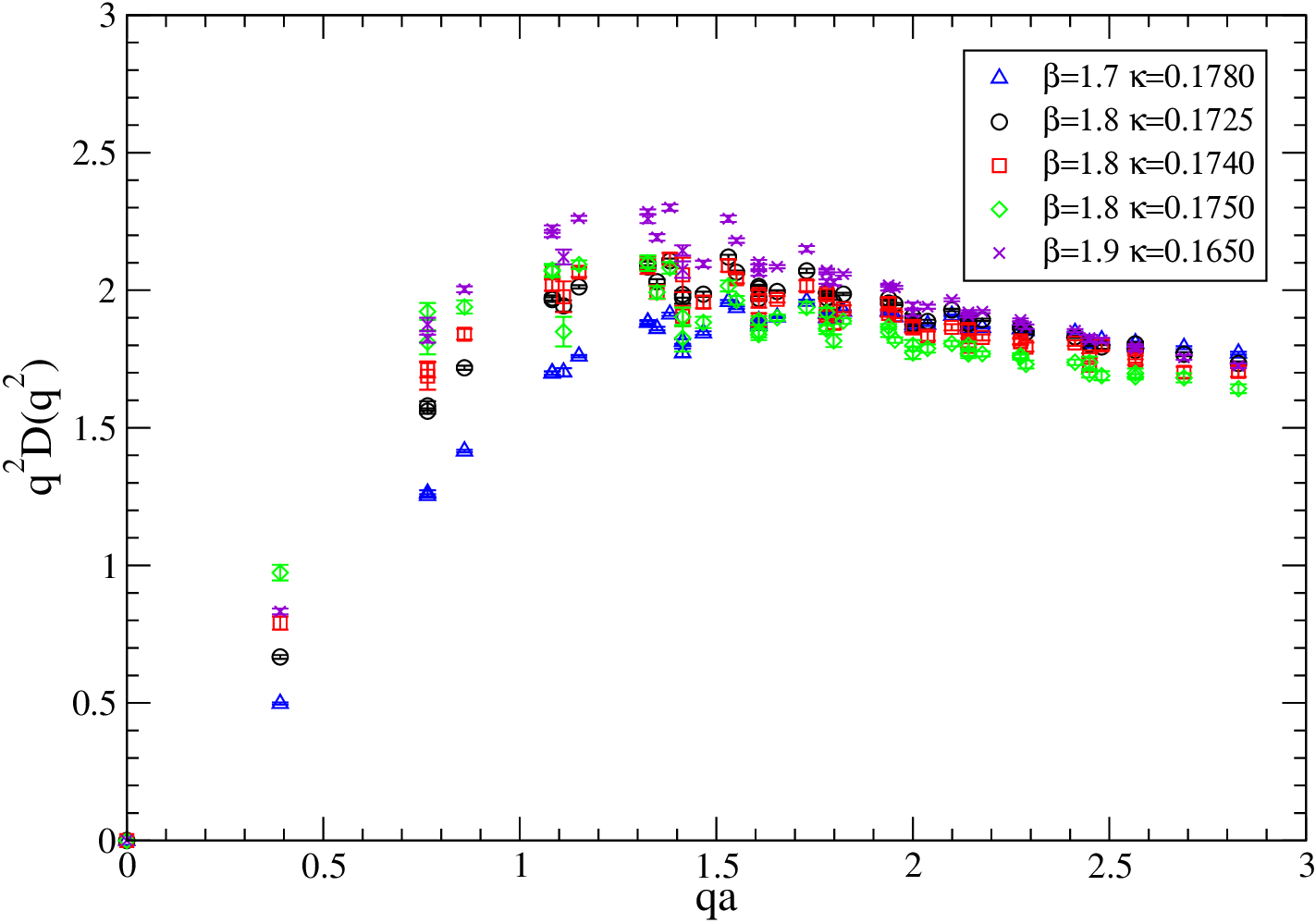}
\includegraphics[width=\halftext,clip]{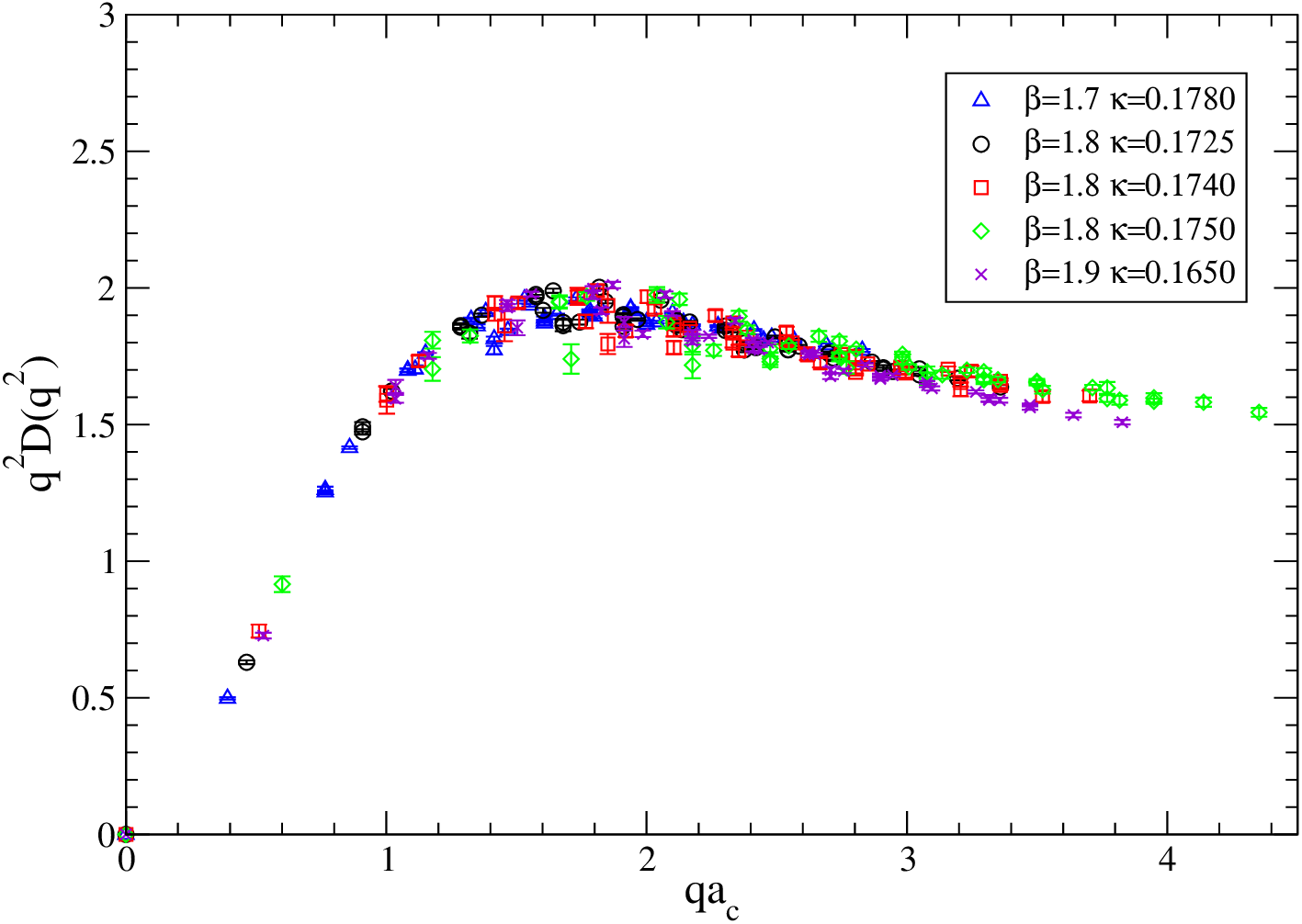}}
\caption{The gluon propagator form factor in Landau gauge, as a
  function of momentum.  Left: before matching of lattice spacings;
  right: after matching.}
\label{fig:gluon}
\end{figure}

We have also attempted to determine the lattice spacing ratios by
matching the (Landau-gauge) gluon propagator as described in
Ref.~\citen{Leinweber:1998uu}.  We have imposed a cylinder cut,
selecting points within a radius $1.2\times\pi/8a$ from the diagonal
in momentum space, and varied the ratios of lattice spacing $a$ and
gluon renormalisation constant $Z_3$ to match the gluon propagator for
the different parameters.  In Fig.~\ref{fig:gluon} we show the gluon
form factor before and after matching.  The resulting $a$ and $Z_3$
ratios are given in table~\ref{tab:Ra}.

This procedure should be treated with some caution, since it can fail
on three counts.  Firstly, in the presence of dynamical fermions the
renormalised gluon propagator will depend on the renormalised fermion
masses, which vary between our parameter sets.  This may explain the
discrepancy between the data for $(\beta=1.8,\kappa=0.175)$ and
$(\beta=1.9,\kappa=0.165)$ at large momenta since, as we shall see,
these two have the lightest and the heaviest quark mass respectively.
Secondly, on these coarse lattices it is likely that we are not in the
scaling regime so the renormalised propagator will have a substantial
cutoff dependence.  Thirdly, since the volume is fixed in lattice
units, these lattices will have different physical volumes and finite
volume effects should cause a slight mismatch in the infrared, where
our procedure has left them almost perfectly matched.  In particular,
for the lightest quark mass, corresponding to the finest lattice, we
see quite large lattice artefacts at intermediate momenta, which may
arise from finite volume effects.  For the other lattices however, the
physical volumes are so large that such effects are likely to be quite
small.

With these caveats in mind, the two methods may complement each other
and give an idea of the systematic uncertainties in the lattice
spacings.  It would clearly be preferable to compute the static quark
potential using smeared Wilson loops, but even this may fail to
provide a reliable lattice spacing on these small lattices.

\subsection{Meson and diquark spectrum}

We have computed correlators, using point sources, for the
pseudoscalar ($\pi$), scalar ($\delta$) and vector ($\rho$) mesons, as
well as for the scalar, pseudoscalar and vector diquarks.  For the
vector diquark we have not been able to extract a mass, while the
scalar and pseudoscalar diquarks are exactly degenerate with the $\pi$
and $\delta$, as they should be at $\mu=0$.  In the following, we will
therefore only quote our results for the meson spectrum, with the main
quantity of interest being $m_\pi/m_\rho$.

\begin{wrapfigure}{r}{\halftext}
\centerline{%
\includegraphics[width=\halftext,clip]{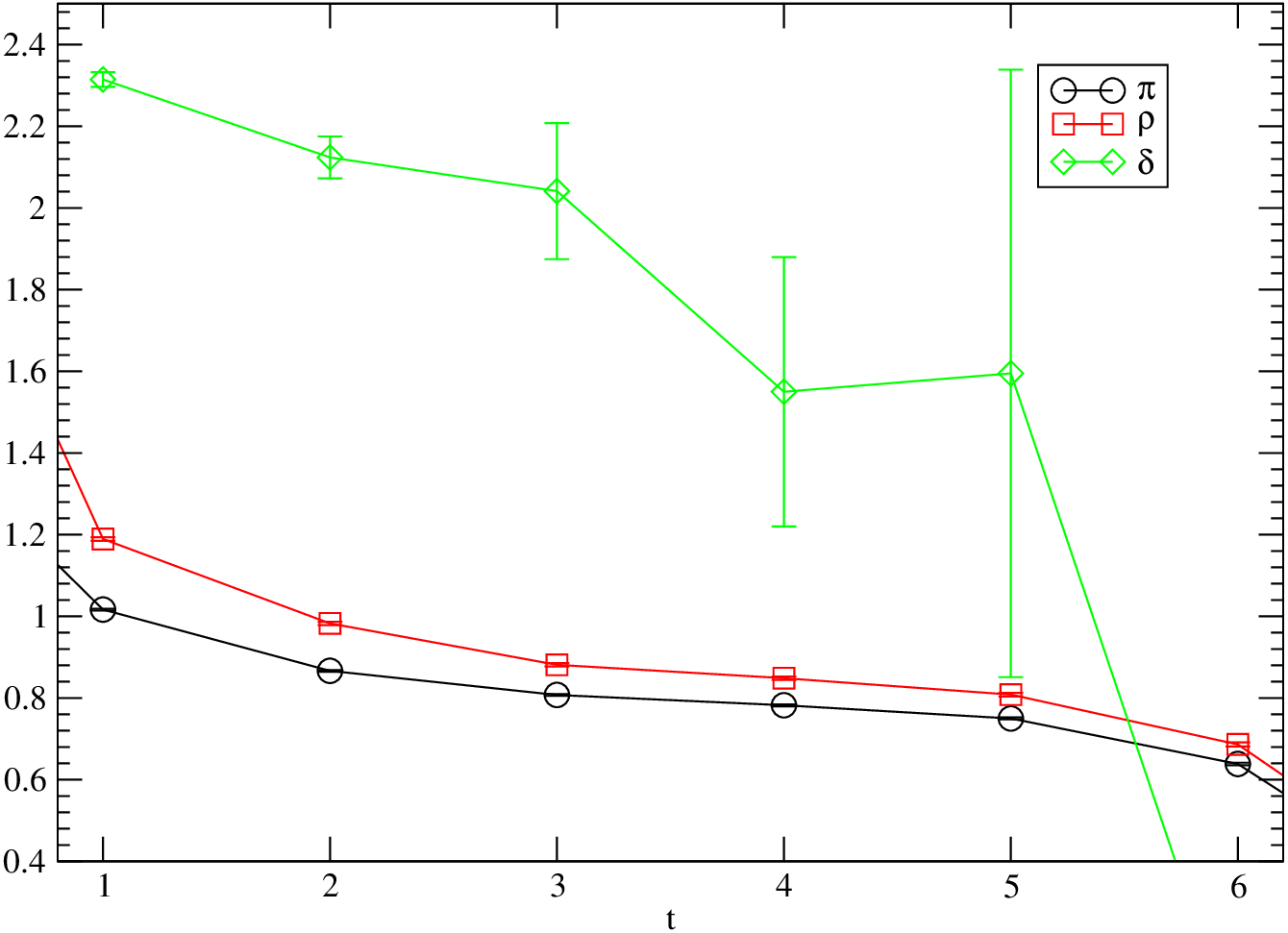}%
}
\caption{Effective masses for $(\beta=0.17,\kappa=0.178)$.}
\label{fig:mesons-b170}
\end{wrapfigure}
In Figure~\ref{fig:mesons-b170} we show the effective meson masses for
$(\beta=1.7,\kappa=0.178)$.  We see, firstly, that with point sources
on these lattices, we do not get a good plateau for either of these
correlators, and secondly, that the $\delta$ correlator is still too
noisy to determine the mass with any confidence, although clearly
$m_\delta\gg m_\pi$, implying we are in a phase where chiral symmetry
is spontaneously broken.
Although we do not see a plateau, we can still fit the $\pi$ and
$\rho$ correlators to a double-exponential (cosh) with a good to
reasonable $\chi^2$.  The resulting masses are given in
table~\ref{tab:masses}.  However, as the absence of a plateau in the
effective-mass plot would indicate, the fit values are not stable, and
one should add a systematic uncertainty of 1--2\% in all quantities
from the variation with the fit range.
\begin{table}
\begin{center}
\begin{tabular}{llllllrl}
$\beta$ & $\kappa$ & $m_\pi$ & $\chi^2/N_{df}$ & $m_\rho$ &
  $\chi^2/N_{df}$ &  range & $m_\pi/m_\rho$ \\ \hline
1.7 &  0.1780 & 0.800(2)  & 0.97  &  0.870(3) & 1.4  & 3--5 & 0.920(3) \\
1.8 &  0.1725 & 0.807(3)  & 1.3   &  0.859(4) & 0.82 & 4--6 & 0.939(4) \\
    &  0.1740 & 0.731(9)  & 2.5   &  0.820(14)& 1.6  & 3--5 & 0.892(15) \\
    &  0.1750 & 0.642(9)  & 0.64  &  0.726(15)& 1.4  & 3--5 & 0.885(16) \\
1.9 &  0.1650 & 0.925(4)  & 0.27  &  0.973(5) & 0.12 & 4--6 & 0.951(4)
\end{tabular}
\end{center}
\caption{Meson masses.  $m_\pi$ and $m_\rho$ are obtained by fitting
  the correlators to a cosh function on the timeslices denoted by
  `range' (and their mirror images around the centre of the lattice).
  The errors are purely statistical bootstrap errors.}
\label{tab:masses}
\end{table}

Comparing tables \ref{tab:Ra} and \ref{tab:masses} it appears that the
lattice spacing changes somewhat more rapidly with $\kappa$ at fixed
$\beta$ than the $\pi$-to-$\rho$ mass ratio.  A more careful analysis
is needed to identify the true lattice spacing and spectrum before it
can be determined whether this is a real effect or merely a sign of
shortcomings in our analysis.

\begin{figure}
\centerline{\includegraphics[width=\halftext,clip]{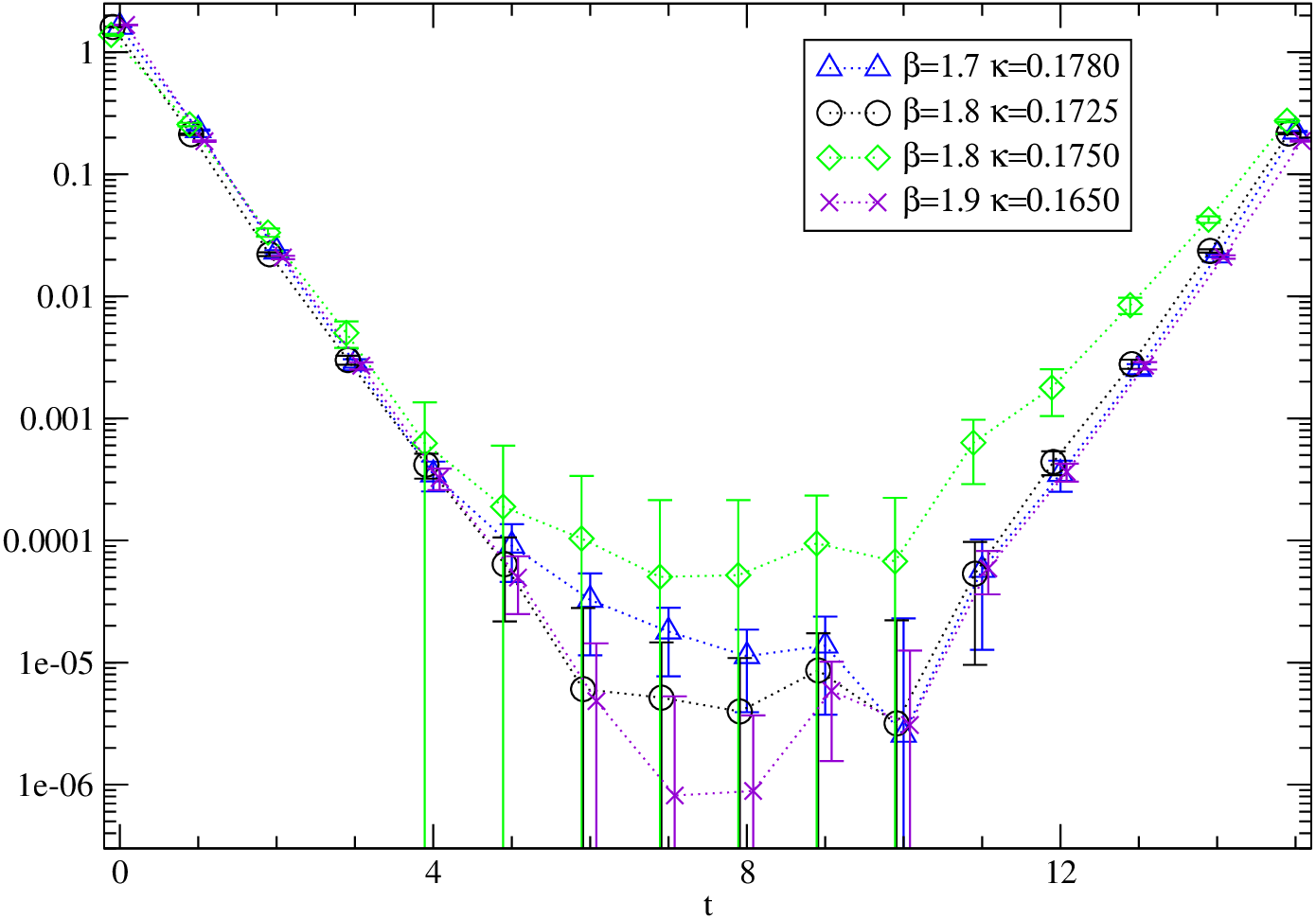}
\includegraphics[width=\halftext,clip]{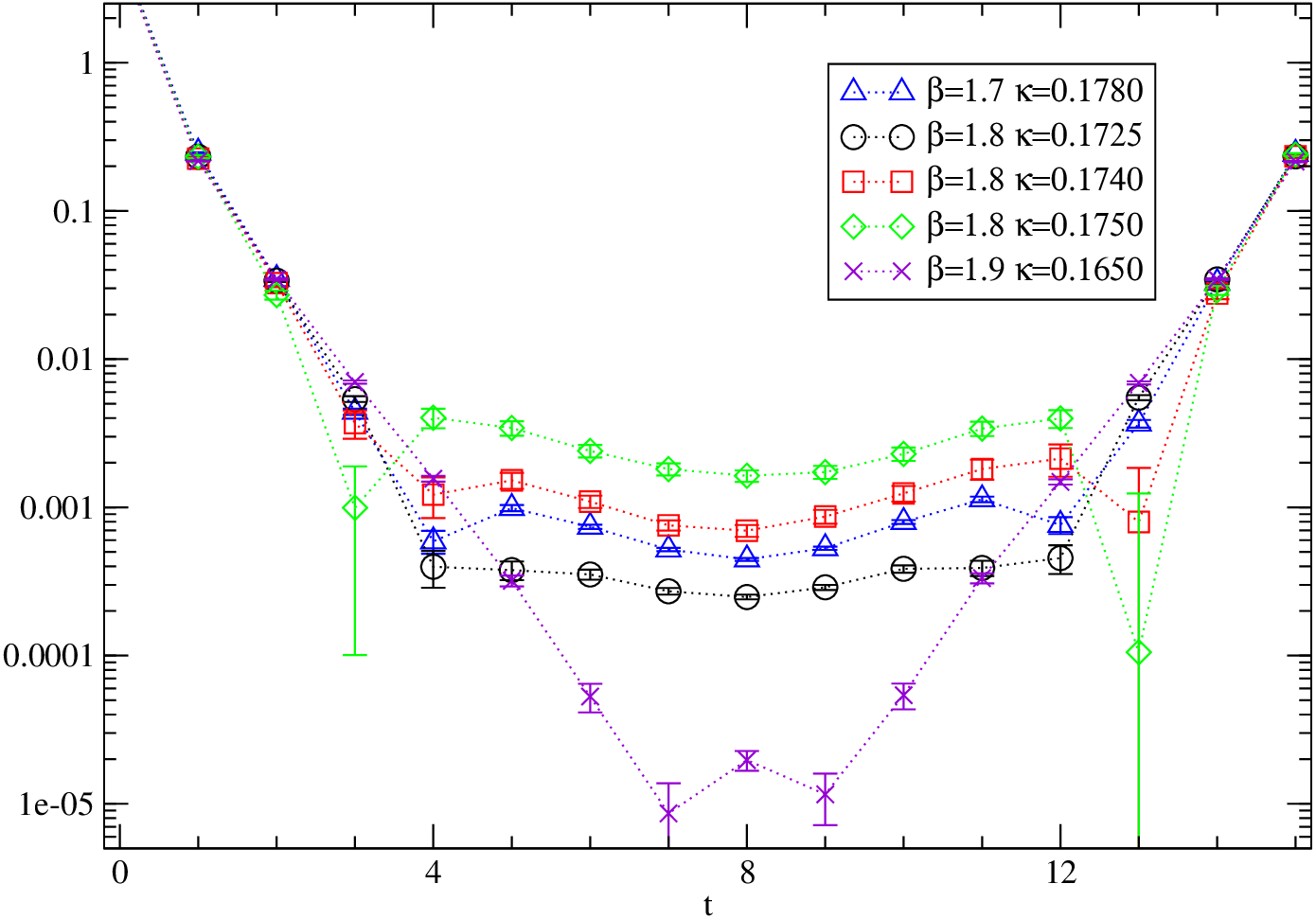}}
\caption{Absolute value of the scalar meson (left) and vector diquark
  (right) correlators.}
\label{fig:delta-dv}
\end{figure}

Finally, in Fig.~\ref{fig:delta-dv} we show the scalar meson and
vector diquark correlators as a function of time.  Both correlators
change sign, so we plot the absolute value in order to show them on a
logarithmic scale.

\section{Outlook}

We have performed initial simulations to explore the parameter space
for 2-colour QCD with Wilson fermions.  The next step will be to
extend the simulations at one or more of these points to non-zero
chemical potential.  Preliminary studies show that HMC experiences a
dramatic slowing-down as we approach the expected transition point at
$\mu=m_\pi/2$.  We will therefore be using the TSMB algorithm to
explore densities beyond this point.

\section*{Acknowledgements}

JIS acknowledges support from FOM/NWO.  SJH is supported by a PPARC
Senior Research Fellowship.

\bibliography{algo,qcd,density,gluon}

\begin{thebibliography}{10}

\bibitem{Muroya:2003qs}
S.~Muroya, A.~Nakamura, C.~Nonaka and T.~Takaishi,
\newblock Prog. Theor. Phys. {\bf 110} (2003), 615 [hep-lat/0306031].

\bibitem{Hands:1999md}
S.~Hands, J.~B. Kogut, M.-P. Lombardo and S.~E. Morrison,
\newblock Nucl. Phys. {\bf B558} (1999), 327 [hep-lat/9902034].

\bibitem{Hands:2000ei}
S.~Hands {\em et~al.},
\newblock Eur. Phys. J. {\bf C17} (2000), 285 [hep-lat/0006018].

\bibitem{Hands:2002mr}
S.~Hands and D.~N. Walters,
\newblock Phys. Lett. {\bf B548} (2002), 196 [hep-lat/0209140].

\bibitem{Walters:2003nn}
D.~N. Walters,
\newblock hep-lat/0310038,
\newblock this volume.

\bibitem{Kogut:2002zg}
J.~B. Kogut and D.~K. Sinclair,
\newblock Phys. Rev. {\bf D66} (2002), 034505 [hep-lat/0202028].

\bibitem{Muroya:2002ry}
S.~Muroya, A.~Nakamura and C.~Nonaka,
\newblock Phys. Lett. {\bf B551} (2003), 305 [hep-lat/0211010].

\bibitem{Muroya:2003zz}
S.~Muroya,
\newblock Color {SU(2)} lattice {QCD} high density state,
\newblock this volume.

\bibitem{Kogut:2000ek}
J.~Kogut, M.~Stephanov, D.~Toublan, J.~Verbaarschot and A.~Zhitnitsky,
\newblock Nucl. Phys. {\bf B582} (2000), 477 [hep-ph/0001171].

\bibitem{Kogut:2002cm}
J.~B. Kogut, D.~Toublan and D.~K. Sinclair,
\newblock Nucl. Phys. {\bf B642} (2002), 181 [hep-lat/0205019].

\bibitem{Brown:1991kk}
G.~E. Brown and M.~Rho,
\newblock Phys. Rev. Lett. {\bf 66} (1991), 2720.

\bibitem{Agakishiev:1998au}
CERES/NA45, G.~Agakishiev {\em et~al.},
\newblock Phys. Lett. {\bf B422} (1998), 405 [nucl-ex/9712008].

\bibitem{Langfeld:1997fi}
K.~Langfeld, H.~Reinhardt and M.~Rho,
\newblock Nucl. Phys. {\bf A622} (1997), 620 [hep-ph/9703342].

\bibitem{Sannino:2001fd}
F.~Sannino and W.~Sch{\"a}fer,
\newblock Phys. Lett. {\bf B527} (2002), 142 [hep-ph/0111098].

\bibitem{Schafer:2002yy}
T.~Sch{\"a}fer,
\newblock Phys. Rev. {\bf D67} (2003), 074502 [hep-lat/0211035].

\bibitem{Hands:2001ee}
S.~Hands, I.~Montvay, L.~Scorzato and J.~Skullerud,
\newblock Eur. Phys. J. {\bf C22} (2001), 451 [hep-lat/0109029].

\bibitem{Roberts:2000aa}
C.~D. Roberts and S.~M. Schmidt,
\newblock Prog. Part. Nucl. Phys. {\bf 45S1} (2000), 1 [nucl-th/0005064].

\bibitem{Alkofer:2000wg}
R.~Alkofer and L.~von Smekal,
\newblock Phys. Rept. {\bf 353} (2001), 281 [hep-ph/0007355].

\bibitem{Lombardo:2003uu}
M.~P. Lombardo, M.~L. Paciello, S.~Petrarca and B.~Taglienti,
\newblock hep-lat/0309110.

\bibitem{Baldo:2003id}
M.~Baldo, P.~Castorina and D.~Zappal{\`a},
\newblock nucl-th/0311038.

\bibitem{Montvay:1996ea}
I.~Montvay,
\newblock Nucl. Phys. {\bf B466} (1996), 259 [hep-lat/9510042].

\bibitem{Leinweber:1998uu}
UKQCD, D.~B. Leinweber, J.~I. Skullerud, A.~G. Williams and C.~Parrinello,
\newblock Phys. Rev. {\bf D60} (1999), 094507 [hep-lat/9811027].

\end{thebibliography}

\end{document}